\documentclass{article}
\usepackage{spconf,amsmath,graphicx, amssymb}
\usepackage{float}
\usepackage{cite}
\usepackage{mathtools}

\usepackage{capt-of}
\usepackage{url}

\title{Attention Is All You Need in Speech Separation}
\name{Cem Subakan$^1$, Mirco Ravanelli$^1$, Samuele Cornell$^2$, Mirko Bronzi$^1$, Jianyuan Zhong$^3$}
\address{$^1$Mila-Quebec AI Institute, Canada,\\
$^2$Università Politecnica delle Marche, Italy \\
$^3$University of Rochester, USA \\
}
\usepackage{tikz}
\usetikzlibrary{shapes,arrows, snakes}

\tikzstyle{specialblock} = [draw, ultra thick, fill=blue!20, rectangle, 
    minimum height=3em, minimum width=4em]
\tikzstyle{block} = [draw, fill=lightgray, rectangle, 
    minimum height=3em, minimum width=4em]
\tikzstyle{sum} = [draw, fill=white, circle, node distance=1cm]
\tikzstyle{prod}   = [circle, minimum width=8pt, draw, inner sep=0pt, path picture={\draw (path picture bounding box.south east) -- (path picture bounding box.north west) (path picture bounding box.south west) -- (path picture bounding box.north east);}]
\tikzstyle{sumt}   = [circle, minimum width=8pt, draw, inner sep=0pt, path picture={\draw (path picture bounding box.east) -- (path picture bounding box.west) (path picture bounding box.south) -- (path picture bounding box.north);}]
\tikzstyle{input} = [coordinate]
\tikzstyle{output} = [coordinate]
\tikzstyle{pinstyle} = [pin edge={to-,thin,black}]
\tikzset{
tmp/.style  = {coordinate}, 
dot/.style = {circle, minimum size=#1,
              inner sep=0pt, outer sep=0pt},
dot/.default = 6pt % size of the circle diameter 
}

\begin{document}
\ninept
\maketitle
\begin{abstract}
Recurrent Neural Networks (RNNs) have long been the dominant architecture in sequence-to-sequence learning. RNNs, however, are inherently sequential models that do not allow parallelization of their computations.  Transformers are emerging as a natural alternative to standard RNNs, replacing recurrent computations with a multi-head attention mechanism.

In this paper, we propose the \textit{SepFormer}, a novel RNN-free Transformer-based neural network for speech separation. The SepFormer learns short and long-term dependencies with a multi-scale approach that employs transformers. The proposed model achieves state-of-the-art (SOTA) performance on the standard WSJ0-2/3mix datasets. It reaches an SI-SNRi of 22.3 dB on WSJ0-2mix and an SI-SNRi of 19.5 dB on WSJ0-3mix. The SepFormer inherits the parallelization advantages of Transformers and achieves a competitive performance even when downsampling the encoded representation by a factor of 8. It is thus significantly faster and it is less memory-demanding than the latest speech separation systems with comparable performance.

\end{abstract}
\begin{keywords}
speech separation, source separation, transformer, attention, deep learning.
\end{keywords}
\section{Introduction}
\label{sec:intro}

RNNs are a crucial component of modern audio processing systems and they are used in many different domains, including speech recognition, synthesis, enhancement, and separation, just to name a few.
Especially when coupled with multiplicative gate mechanisms (like LSTM \cite{lstm} and GRU \cite{gru,li_gru}), their recurrent connections are essential to learn long-term dependencies and properly manage speech contexts. 
Nevertheless, the inherently sequential nature of RNNs impairs an effective parallelization of the computations.  This bottleneck is particularly evident when processing large datasets with long sequences. 
On the other hand, Transformers \cite{vaswani2017} completely avoid this bottleneck by eliminating recurrence and replacing it with a fully attention-based mechanism. By attending to the whole sequence at once, a direct connection can be established between distant elements allowing Transformers to learn long-term dependencies more easily \cite{kerg2020untangling}.
For this reason, Transformers are gaining considerable popularity for speech processing and recently showed competitive performance in speech recognition \cite{transformers_espnet}, synthesis \cite{transformers_tts}, enhancement \cite{transformers_se}, diarization \cite{transformers_diarization}, as well as speaker recognition \cite{transformers_speaker}. 

Little research has been done so far on Transformer-based models for monaural audio source separation. The field has been revolutionized by the adoption of deep learning techniques \cite{hershey2015deep,  yu2017permutation, kolbaek2017multitalker, venkataramani2017endtoend, luo2018convtasnet, Huang2014}, and with recent works \cite{luo2020dualpath, liu2019divide, shi2019furcanext, nachmani2020voice, tzinis2020sudo, dptn, zeghidour2020wavesplit} achieving impressive results by adopting an end-to-end approach. %Except for  Wavesplit \cite{zeghidour2020wavesplit} that uses speaker identity to drive separation, 
Most of the current speech separation techniques \cite{venkataramani2017endtoend, luo2018convtasnet, liu2019divide, luo2020dualpath, shi2019furcanext, nachmani2020voice, tzinis2020sudo,  dptn} require effective modeling of long input sequences to perform well.
Current systems rely, in large part, on the learned-domain masking strategy popularized by Conv-TasNet \cite{luo2018convtasnet}. In this framework, an overcomplete set of analysis and synthesis filters is learned directly from the data, and separation is performed by estimating a mask for each source in this learned-domain.
Building on this, Dual-Path RNN (DPRNN) \cite{luo2020dualpath} has demonstrated that better long-term modeling is crucial to improve the separation performance. This is achieved by splitting the input sequence into multiple chunks that are processed locally and globally with different RNNs. Nevertheless, due to the use of RNNs, DPRNN still suffers from the aforementioned limitations of recurrent connections, especially regarding the global processing step. 
An attempt to integrate transformers into the speech separation pipeline has been recently done in \cite{dptn} where the proposed Dual-Path Transformer Network (DPTNet) is shown to outperform the standard DPRNN. Such an architecture, however, still embeds an RNN, effectively negating the parallelization capability of pure-attention models. 

\newcommand{\mixture}{x}
\newcommand{\ldim}{F}
\newcommand{\llen}{T'}
\newcommand{\nspk}{Ns}
\newcommand{\nsepf}{M}
\newcommand{\chnksize}{C}
\newcommand{\hopsize}{H}
\newcommand{\numchnks}{Nc}
\newcommand{\numintra}{Nintra}
\newcommand{\numinter}{Ninter}

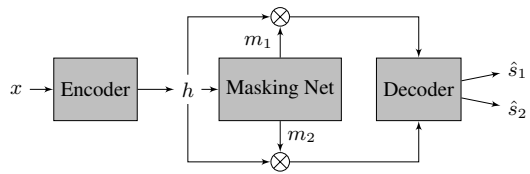
\begin{figure}[t!]
\centering
    \resizebox{7.2cm}{!}{
    \begin{tikzpicture}[auto, node distance=2.0cm,>=latex']
        \node [draw=none, fill=none] (input) {$\mixture$ };
        \node [block, right of=input, fill, xshift=-0.8cm ] (encoder) {Encoder};  
        \node [draw=none, fill=none, right of=encoder, xshift=-0.6cm] (latent) {$h$};
        \node [block, right of=latent, xshift=-0.6cm] (masknet) {Masking Net};
        \node [prod, above of=masknet,yshift=-0.9cm] (prod1) {}; 
        \node [prod, below of=masknet,yshift=0.9cm] (prod2) {}; 
        \node [block, right of=masknet, xshift=0.1cm] (decoder) {Decoder};
        \node [draw=none, fill=none, right of=decoder, yshift=0.3cm, xshift=-0.5cm] (out1) {$\hat{s}_1$};
        \node [draw=none, fill=none, right of=decoder, yshift=-0.3cm, xshift=-0.5cm] (out2) {$\hat{s}_2$};
        %\node [draw=none, fill=none, right of=masknet, yshift=0.1cm, xshift=-0.5cm] (mid1) {};
        %\node [draw=none, fill=none, right of=masknet, yshift=-0.1cm, xshift=-0.5cm] (mid2) {};
        
        \draw [->] (input) -- (encoder);
        \draw [->] (encoder) -- (latent);
        \draw [->] (latent) -- (masknet);
        \draw [->] (latent) |-  (prod1);
        \draw [->] (latent) |-  (prod2);
        \draw [->] (masknet) -- node {$m_1$} (prod1);
        \draw [->] (masknet) -- node {$m_2$} (prod2);
        \draw [->] (prod1) -| (decoder);
        \draw [->] (prod2) -| (decoder);
        %\draw [->] (mid1) -- (decoder);
        %\draw [->] (mid2) -- (decoder);
        \draw [->] (decoder) -- (out1);
        \draw [->] (decoder) -- (out2);
    \end{tikzpicture}
    }
\caption{The high-level description of our system: The encoder block estimates a learned-representation for the input signal, while the masking network estimates optimal masks to separate the sources present in the mixtures. The decoder finally reconstructs the estimated sources in the time domain using the masks provided by the masking network. }
\label{fig:tasnetpipeline}
\end{figure}

In this paper, we propose a novel model called SepFormer (Separation Transformer), which is mainly composed of multi-head attention and feed-forward layers. We adopt the dual-path framework introduced by DPRNN and we replace RNNs with a multi-scale pipeline composed of transformers that learn both short and long-term dependencies. The dual-path framework enables to mitigate the quadratic complexity of transformers, as transformers in the dual-path framework process smaller chunks.  

%The transformers are not directly applicable to long audio signals because of its quadratic complexity in sequence length. However, the dual-path framework enables the usage of transformers in audio-source separation as it divides the signal into smaller chunks, and processes the chunks independently. 

To the best of our knowledge, this is the first work showing that we can obtain state-of-the-art performance in separation with an RNN-free Transformer-based architecture. The SepFormer achieves an SI-SNRi of 22.3 dB on the standard WSJ0-2mix dataset. It also achieves the SOTA performance of 19.5 dB SI-SNRi on the WSJ0-3mix dataset. The SepFormer not only processes all the time steps in parallel but also achieves competitive performance when downsampling the encoded representation by a factor of 8. This makes the proposed architecture significantly faster and less memory demanding than the latest RNN-based separation models.

\begin{figure*}[t!]
\centering
    \newcommand{\sepfig}{0.5}
    \begin{tikzpicture}[auto, node distance=2.0cm,>=latex']
        \node [draw=none, fill=none] (h) {$h$};
        \node [block, right of=h, fill, xshift=-0.3cm] (1) {Norm+Linear};  
        \node [block, right of=1, fill, xshift=\sepfig cm] (2) {Chunking};  
        \node [specialblock, right of=2,  xshift=0.3 cm] (3) {\textbf{SepFormer}};  
        \node [block, right of=3, xshift=\sepfig cm] (4) {PReLU+Linear};
        \node [block, right of=4, xshift=0.6 cm] (5) {OverlapAdd};
        \node [block, right of=5, xshift=0.6 cm] (6) {FFW+ReLU};
        %\node [block, right of=2, fill, xshift=1cm] (4) {\textbf{Sepformer}};  
        \node [draw=none, fill=none, right of=6, yshift=0.3cm] (out1) {$m_1$};
        \node [draw=none, fill=none, right of=6, yshift=-0.3cm] (out2) {$m_2$};
        
        \draw [->] (h) -- (1);
        \draw [->] (1) -- (2); 
        \draw [->] (2) -- node {$h'$} (3); 
        \draw [->] (3) -- node {$h''$} (4); 
        \draw [->] (4) -- node {$h'''$} (5); 
        \draw [->] (5) -- node {$h''''$} (6); 
        \draw [->] (6) -- (out1); 
        \draw [->] (6) -- (out2); 
    \end{tikzpicture}
    \vspace{0.2cm}
    
     \resizebox{10cm}{!}{    
    \begin{tikzpicture}[auto, node distance=2cm,>=latex']
        \node [draw, ultra thick, fill=blue!20, rectangle, right of=1, xshift=1.0cm, minimum width= 8.0cm, minimum height= 1.5cm] (plate) {};
        \node [draw=none, below of=plate,yshift=1cm, xshift=3.0cm] (platetext) {\textit{Repeat N times}};
        \node [draw=none, fill=none, xshift=0cm] (input) {$h'$};
        \node [block, right of=input, xshift=0.5cm] (1) {IntraTransformer};  
        \node [block, right of=1, xshift=0.2cm] (2) {Permute};
        \node [block, right of=2, xshift=0.2cm] (3) {InterTransformer};  
        \node [draw=none, right of=3, xshift=0.5cm] (out) {$h''$};
        
        \draw [->] (input) -- (1);
        \draw [->] (1) -- (2);
        \draw [->] (2) -- (3); 
        \draw [->] (3) -- (out);
    \end{tikzpicture}
    }
    
    \vspace{0.0cm}
   \resizebox{10.8cm}{!}{
    \begin{tikzpicture}[auto, node distance=2cm,>=latex']
        \node [draw, thick, fill=gray!20, rectangle, right of=1, xshift=2.1cm, minimum width= 10.9cm, minimum height= 1.9cm] (plate) {};
        \node [draw=none, below of=plate,yshift=0.8cm, xshift=4.3cm] (platetext) {\textit{Repeat K times}};
        \node [draw=none, ] (input) {$z$};
        \node [sumt, right of=input, xshift=-1.2cm] (sum) {};
        \node [draw=none, above of=sum, yshift=-1.3cm] (e) {$e$};
        \node [block, right of=sum] (1) {LayerNorm};
        \node [block, right of=2, xshift=-1.8cm] (2) {MHA};
        \node [sumt, right of=2, xshift=-0.4cm] (sum2) {};
        \node [block, right of=sum2, xshift=-0.3cm] (3) {LayerNorm};
        \node [block, right of=3] (4) {FFW};
        \node [sumt, right of=4, xshift=-0.6cm] (sum3) {};
        \node [draw=none, right of=sum3, xshift=-0.7cm] (out) {$z'''$};
        \node [sumt, right of=sum3, xshift=0.1cm] (sum4) {};
        \node [draw=none, right of=sum4, xshift=-1.0cm] (fout) {$f(z)$};
        \node [tmp, below of=1, yshift=1.3cm] (tmp1) {};
        \node [tmp, below of=sum2, yshift=1.3cm] (tmp2) {};
        
        \draw [->] (input) -- (sum);
        \draw [->] (e) -- (sum);
        \draw [->] (sum) -- node (zp) {$z'$} (1);
        \draw [->] (1) -- (2);
        \draw [->] (2) -- node (zpp) {$z''$}(sum2); 
        \draw [->] (sum2) -- node (zinp) {} (3);
        \draw [->] (3) -- (4);
        \draw [->] (4) -- (sum3);
        \draw [->] (sum3) -- (out);
        \draw [->] (zp) |-(tmp1) -| (sum2);
        %\draw [->] (tmp2) -| (sum3);
        \draw [->] (out) -- (sum4);
        \draw [->] (sum4) -- (fout);
        
        \node [tmp, below of=zinp, yshift=1.18cm] (tmp4) {};
        % for the missing residual connection
        \node [tmp, below of=zp, yshift=.4cm] (tmp3) {};
        \draw [-] (input) |- (tmp3);
        \draw[->] (tmp3) -| (sum4);
        \draw [->] (zinp) |- (tmp4) -| (sum3);
        
    \end{tikzpicture}
    }    
    \vspace{-0.3cm}
    \caption{\textbf{(Top)} The overall architecture proposed for the masking network. \textbf{(Middle)} The SepFormer Block.  \textbf{(Bottom)} The transformer architecture $f(.)$ that is used both in the IntraTransformer block and in the InterTransformer block.}
\label{fig:sepformer}
\end{figure*}
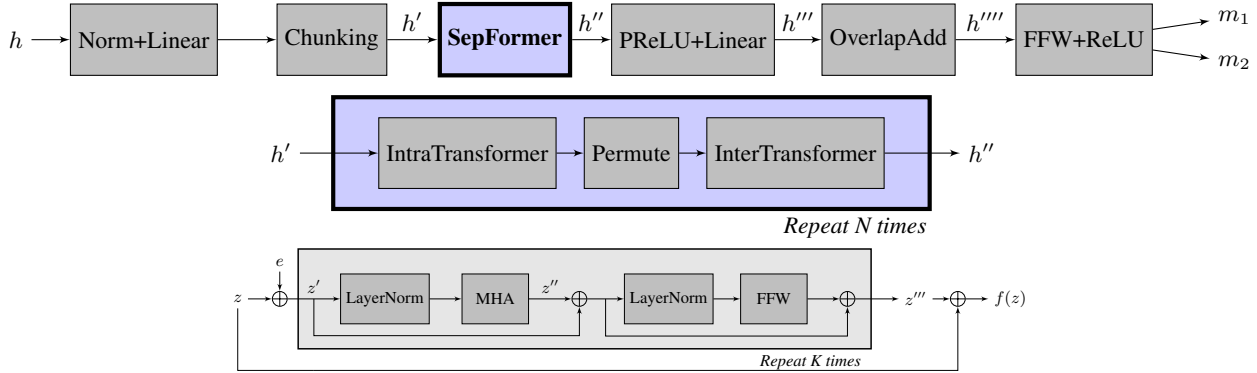

\section{The model}
The proposed model is based on the learned-domain masking approach \cite{venkataramani2017endtoend, luo2018convtasnet, liu2019divide, luo2020dualpath, shi2019furcanext, nachmani2020voice, tzinis2020sudo,  dptn} and employs an encoder, a decoder, and a masking network, as shown in Figure \ref{fig:tasnetpipeline}.
The encoder is fully convolutional, while the masking network employs two Transformers embedded inside the dual-path processing block proposed in \cite{luo2020dualpath}. The decoder finally reconstructs the separated signals in the time domain by using the masks predicted by the masking network. To foster reproducibility, the SepFormer will be made available within the SpeechBrain toolkit\footnote{\url{speechbrain.github.io/}}.

\subsection{Encoder}
The encoder takes in the time-domain  mixture-signal $\mixture \in \mathbb R^T$ as input, which contains audio from multiple speakers. 
It learns an STFT-like representation $h \in \mathbb R^{\ldim \times \llen}$ using a single convolutional layer:
\begin{align}
    h = \text{ReLU}(\text{conv1d}(x)).  
\end{align}
As we will describe in Sec. \ref{sec:res}, the stride factor of this convolution impacts significantly on the performance, speed, and memory of the model.

\subsection{Masking Network}
Figure \ref{fig:sepformer} (top) shows the detailed architecture of the masking network (Masking Net).
The masking network  is fed by the encoded representations $h \in \mathbb R^{\ldim \times \llen}$ and estimates a mask $\{m_1,\dots,m_{\nspk}\}$ for each of the $\nspk$ speakers in the mixture. 

As in \cite{luo2018convtasnet}, the encoded input $h$ is normalized with layer normalization \cite{layernorm} and processed by a linear layer (with dimensionality $\ldim$). 
We then create overlapping chunks of size $\chnksize$ by chopping up $h$ on the time axis with an overlap factor of 50\%. We denote the output of the chunking operation with $h' \in \mathbb R^{\ldim \times \chnksize \times \numchnks}$, where $C$ is the length of each chunk, and $\numchnks$ is the resulting number of chunks. 

The representation $h'$ feeds the SepFormer block, which is the main component of the masking network. This block, which will be described in detail in Sec. \ref{sec:sepformer}, employs a pipeline composed of two transformers able to learn short and long-term dependencies.

The output of the SepFormer $h'' \in \mathbb R^{\ldim \times \chnksize \times \numchnks}$ is processed by PReLU activations followed by a linear layer.  
We denote the output of this module $h''' \in \mathbb R^{(\ldim \times \nspk) \times \chnksize \times\numchnks}$, where $\nspk$ is the number of speakers. Afterwards we apply the overlap-add scheme described in \cite{luo2020dualpath} and obtain $h''''\in \mathbb R^{\ldim \times \nspk \times T'}$. We pass this representation through two feed-forward layers and a ReLU activation at the end to finally obtain the mask $m_k$ for each of the speakers.
\vspace{-0.3cm}
\subsection{SepFormer Block}
\label{sec:sepformer}
Figure \ref{fig:sepformer} (Middle) shows the architecture of the SepFormer block. 
The SepFormer block is designed to model both short and long-term dependencies with the dual-scale approach of DPRNNs \cite{luo2020dualpath}. In our model, the transformer block which models the short-term dependencies is named IntraTransformer (IntraT), and the block for longer-term dependencies is named InterTransformer (InterT). IntraT processes the second dimension of $h'$, and thus acts on each chunk independently, modeling the short-term dependencies within each chunk.
Next, we permute the last two dimensions (which we denote with $\mathcal{P}$), and the InterT is applied to model the transitions across chunks. This scheme enables effective modelling of long-term dependencies across the chunks. The overall transformation of the SepFormer is therefore defined as follows:
\begin{align}
    h'' = f_\text{inter}( \mathcal P  ( f_\text{intra} ( h')) ), 
\end{align}
where we denote the IntraT and InterT with $f_\text{inter}(.)$, and $f_\text{intra}(.)$, respectively.
The overall SepFormer block %consisting of IntraT and InterT 
is repeated $N$ times. 

\subsubsection{Intra and Inter Transformers}
Figure \ref{fig:sepformer} (Bottom) shows the  architecture of the Transformers used for both the IntraT and InterT blocks. It closely resembles the original one defined in \cite{vaswani2017}.
We use the variable $z$ to denote the input to the Transformer. 
First of all, sinusoidal positional encoding $e$ is added to the input $z$, such that, 
\begin{align}
    z' = z + e.
\end{align}
Positional encoding injects information on the order of the various elements composing the sequence, thus improving the separation performance. We follow the positional encoding definition in \cite{vaswani2017}. 

We then apply multiple Transformer layers. %which involve normalizations,
Inside each Transformer layer $g(.)$, we first apply layer normalization, followed by multi-head attention (MHA):
\begin{align}
    z''= \text{MultiHeadAttention}(\text{LayerNorm}(z')).
\end{align}
As proposed in \cite{vaswani2017}, each attention head computes the scaled dot-product attention between all the elements of the sequence.
The Transformer finally employs a feed-forward network (FFW), which is applied to each position independently:
\begin{align}
    z''' = \text{FeedForward}(\text{LayerNorm}(z''+ z') ) + z'' + z'.
    \label{eq_ff}
\end{align}
The overall transformer block is therefore defined as follows:
\begin{align}
    f(z) =  g^K (z+e) + z, 
    \label{eq_intra}
\end{align}
where $g^K(.)$ denotes $K$ layers of transformer layer $g(.)$. We use $K=\numintra$ layers for the IntraT, and $K=\numinter$ layers for the InterT.
As shown in Figure \ref{fig:sepformer} (Bottom) and Eq. \eqref{eq_intra}, we add residual connections across the transformer layers, and across the transformer architecture to improve gradient backpropagation. 

\vspace{-0.2cm}
\subsection{Decoder}
The decoder simply uses a transposed convolution layer, with the same stride and kernel size of the encoder. The input to the decoder is the element-wise multiplication between the mask $m_k$ of the source $k$ and the output of the encoder $h$. The transformation of the decoder can therefore be expressed as follows: 
\begin{align}
    \widehat {s}_k = \text{conv1d-transpose}(m_k * h), 
\end{align}
\vspace{-0.3cm}
where $\widehat{s}_k \in \mathbb R^T$ denotes the separated source $k$. 

\section{Experimental Setup}
\vspace{-0.2cm}
\subsection{Dataset}
We use the popular WSJ0-2mix and WSJ0-3mix datasets \cite{hershey2015deep} for source separation, where mixtures of two speakers and three speakers are created by randomly mixing utterances in the WSJ0 corpus. The relative levels for the sources are sampled uniformly between 0 dB to 5 dB. Respectively, 30, 10, 5 hours of speech is used for training, validation, and test. The training and test sets are created with different sets of speakers. The waveforms are sampled at 8 kHz.

% short description of the corpora used
\vspace{-0.2cm}
\subsection{Architecture and Training Details}
\label{sec:arcdetails}
The encoder is based on 256 convolutional filters with a kernel size of 16 samples and a stride factor of 8 samples. The decoder uses the same kernel size and the stride factors of the encoder. 

%Before feeding the mixtures into the encoder, we apply waveform dropout by randomly zeroing small random chunks of consecutive samples. We also randomly drop out some frequency channels by convolving the waveform with random band-pass filters as done in \cite{pase}.

In our best models, the SepFormer masking network processes chunks of size $\chnksize=250$ with a 50 \% overlap between them and employs 8 layers of transformers in both IntraT and InterT. The IntraT-InterT dual-path processing pipeline is repeated $N=2$ times. We used 8 parallel attention heads, and 1024-dimensional positional feed-forward networks within each Transformer layer. The model has a total of 26 million parameters.

\newcommand{\dataaug}{DM}
%Training is performed with the WSJ0-2mix or  WSJ0-3mix data without using on-the-fly mixture generation to create unseen speaker combinations. Data contamination with noise or reverberation \cite{contamiation} is not used as well.
We explored the use of dynamic mixing (DM) data augmentation \cite{zeghidour2020wavesplit} which consists in on-the-fly creation of new mixtures from single speaker sources. In this work we expanded this powerful technique by applying also speed perturbation on the sources before mixing them. The speed randomly changes between 95 \% slow-down and 105 \% speed-up.
%We used randomly sampled independent speed changes for each source.

%(\dataaug=2 in Table \ref{tab:ablation}), and finally in addition to the speed change we also explored creating mixtures by randomly mixing the sources (\dataaug=3 in Table \ref{tab:ablation}) \cite{zeghidour2020wavesplit}. We refer to this option as dynamic mixing (DM) in Tables \ref{table:WSJ2Mix}, \ref{table:3mix}.  

We used the Adam algorithm \cite{kingma2017adam} as optimizer, with a learning rate of $15e^{-5}$. After epoch 65 (after epoch 100 with \dataaug), the learning rate is annealed by halving it if we do not observe any improvement of the validation performance for 3 successive epochs (5 epoch for \dataaug). Gradient clipping is employed to limit the $L2$ norm of the gradients to 5. During training, we used a batch size of 1, and used the scale-invariant signal-to-noise Ratio (SI-SNR) \cite{le2019sdr} via utterance-level permutation invariant loss \cite{kolbaek2017multitalker}, with clipping at 30dB \cite{zeghidour2020wavesplit}. We used automatic mixed-precision to speed up training. The system is trained for a maximum of 200 epochs. Each epoch takes approximately 1.5 hours on a single NVIDIA V100 GPU with 32 GB of memory.  

\vspace{-0.3cm}
\section{Results}
\label{sec:res}
\begin{table}[t]
\centering
\caption{Best results on the WSJ0-2mix dataset (test-set). DM stands for dynamic mixing. } 
\vspace{0.2cm}
\label{table:WSJ2Mix}
\resizebox{8cm}{!}{
\begin{tabular}{l|c|c|c|c}
\textbf{Model} & \textbf{SI-SNRi} & \textbf{SDRi} & \textbf{\# Param} & \textbf{Stride } \\
\hline \hline 
Tasnet \cite{luo2017tasnet} & 10.8 & 11.1 & n.a & 20  \\ \hline
SignPredictionNet \cite{wang2018deep} & 15.3 & 15.6 & 55.2M & 8\\ \hline
ConvTasnet \cite{luo2018convtasnet} & 15.3  & 15.6 & 5.1M & 10\\ \hline 
Two-Step CTN \cite{tzinis2019twostep} & 16.1 & n.a. & 8.6M & 10 \\ \hline
DeepCASA \cite{liu2019divide} & 17.7 & 18.0 & 12.8M & 1\\ \hline
FurcaNeXt \cite{shi2019furcanext} & n.a. & 18.4 & 51.4M & n.a.\\ \hline 
DualPathRNN \cite{luo2020dualpath} & 18.8 & 19.0 & 2.6M & 1 \\ \hline
sudo rm -rf \cite{tzinis2020sudo} &  18.9 & n.a. & 2.6M  & 10\\ \hline 
VSUNOS \cite{nachmani2020voice} & 20.1 & 20.4 & 7.5M & 2 \\ \hline
DPTNet* \cite{dptn} & 20.2 & 20.6 & 2.6M & 1\\  \hline 
Wavesplit** \cite{zeghidour2020wavesplit} & 21.0 & 21.2 & 29M & 1 \\ 
Wavesplit** + DM \cite{zeghidour2020wavesplit} & 22.2 & 22.3 & 29M & 1 \\ \hline
\hline
\textbf{SepFormer} & 20.4 & 20.5 & 26M & 8 \\ 
%\textbf{SepFormer + Speed} &  21.8 & 21.9 & 26M & 8 \\
\textbf{SepFormer + DM} &  22.3 & 22.4 & 26M & 8 \\
\hline \hline
\end{tabular}
}
\vspace{1ex}
{\raggedright \footnotesize{*only SI-SNR and SDR (without improvement) are reported.} \par}
\vspace{-1ex}
{\raggedright \footnotesize{**uses speaker-ids as additional info.} \par}
\end{table}
\vspace{-0.2cm}

\subsection{Results on WSJ0-2mix}
Table \ref{table:WSJ2Mix} compares the performance achieved by the proposed SepFormer with the best results reported in the literature on the WSJ0-2mix dataset. The SepFormer achieves an SI-SNR improvement (SI-SNRi) of 22.3 dB and a Signal-to-Distortion Ratio \cite{vincent2006performance} (SDRi) improvement of 22.4 dB on the test-set with dynamic mixing. When using  dynamic mixing, the proposed architecture achieves state-of-the-art performance. The SepFormer outperforms previous systems without using dynamic mixing except Wavesplit, which uses speaker identity as additional information. 
%Note that DPTNet only reports results in terms of absolute SI-SNR and SDR values (without improvement). According to our calculation, SDR is greater than SDRi of about 0.15 dB on the WSJ0-2mix dataset. We do not outperform Wavesplit\cite{zeghidour2020wavesplit} on this dataset, which uses speaker identity to drive separation as an additional information.

\begin{figure*}[t!]
    \centering
    \includegraphics[width=0.292\textwidth]{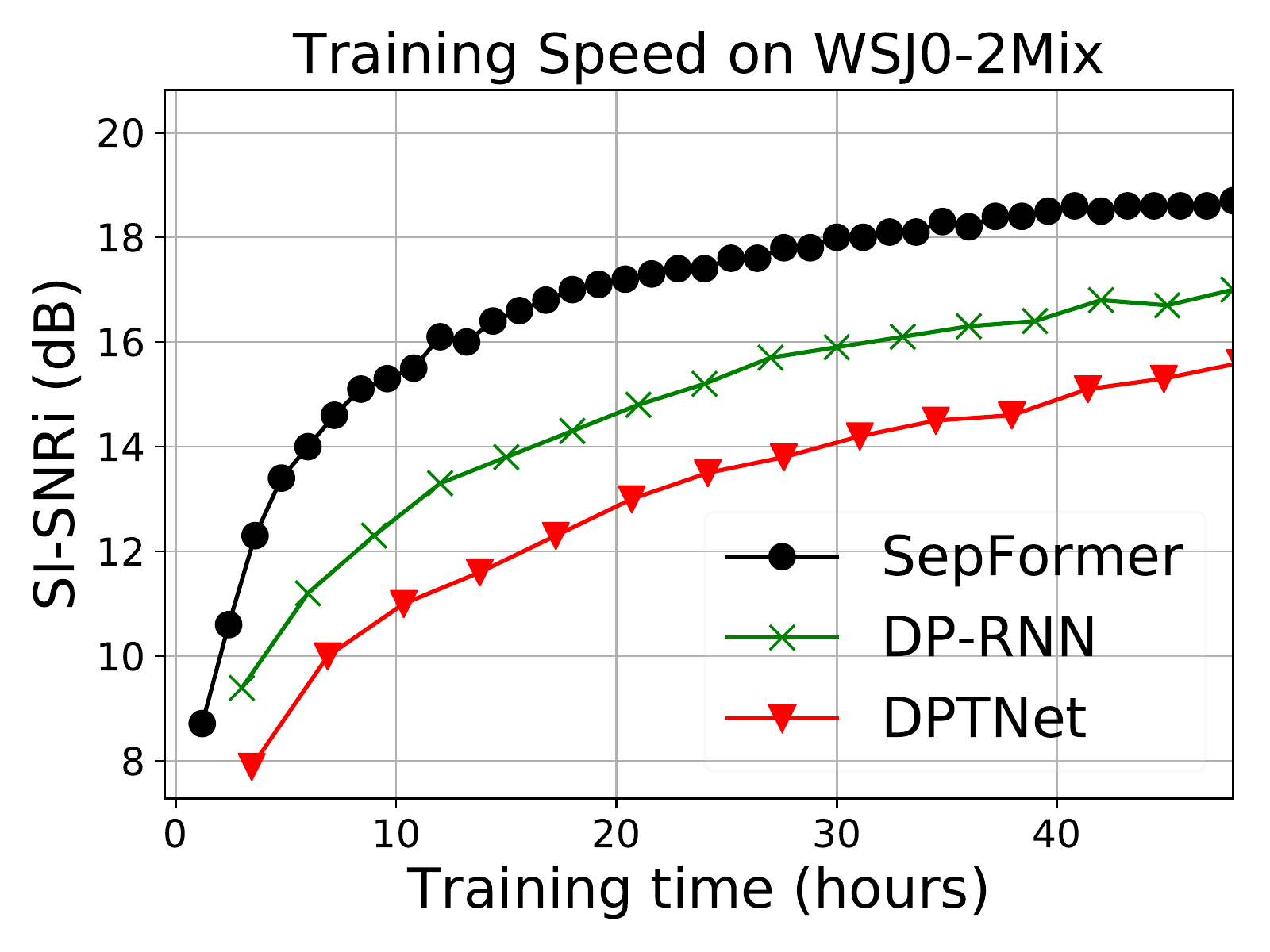}
    \includegraphics[width=.64\textwidth, trim=0cm 0cm 0cm 0cm, clip]{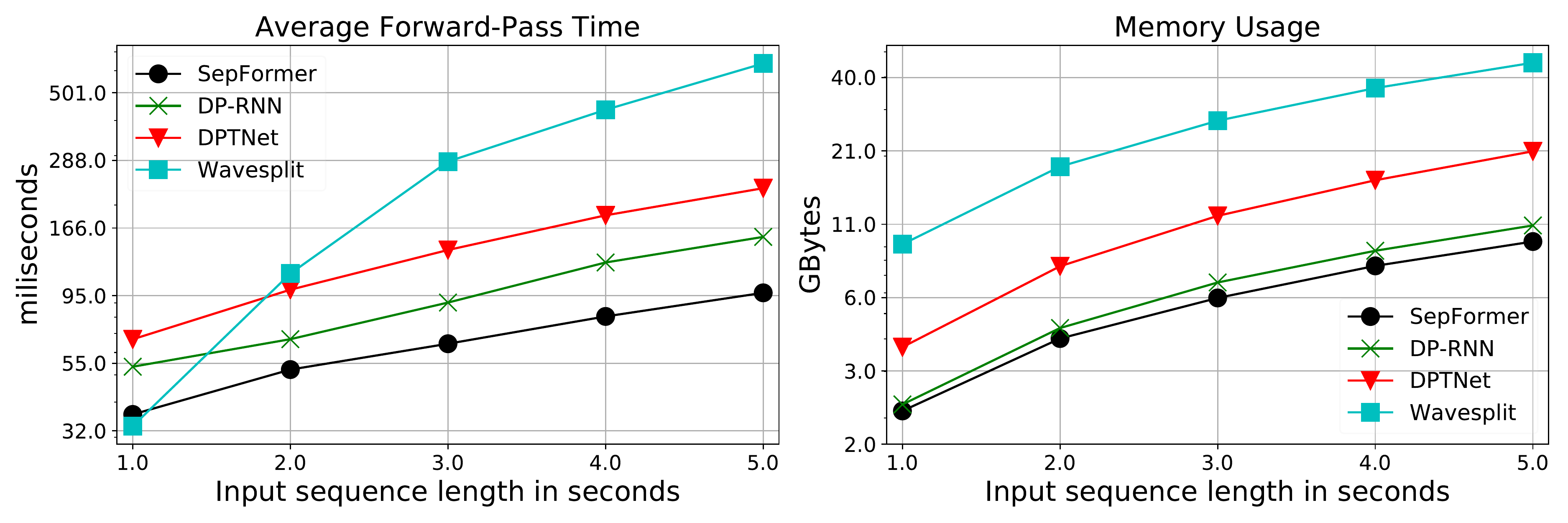}
    \vspace{-0.2cm}
    \caption{(\textbf{Left}) The traning curves of SepFormer, DPRNN, and DPTNeT on the WSJ0-2mix dataset. \textbf{(Middle \& Right)} The comparison of forward-pass speed and memory usage in the GPU %for SepFormer, DPRNN, and DPTNeT 
    on inputs ranging 1-5 seconds long sampled at 8kHz.}
    \label{fig:forwardpass}
    \vspace{-0.2cm}
\end{figure*}

\vspace{-0.2cm}
\begin{table}[t!]
    \caption{Ablation of the SepFormer on WSJ0-2Mix (validation set).}
    \vspace{0.1cm}
    \resizebox{8.6cm}{!}{
    \begin{tabular}{c|c|c | c| c | c | c | c  }
        \textbf{SI-SNRi} & \textbf{$N$} & $\numintra$ & $\numinter$ & \textbf{ \# Heads} & \textbf{DFF} &  \textbf{PosEnc} & \textbf{\dataaug}  \\ \hline \hline
        {22.3} & {2} & {8} & {8} & {8} & {1024} & {Yes} & Yes \\ \hline 
        %{21.8} & {2} & {8} & {8} & {8} & {1024} & {Yes} & 2 \\ \hline 
        %{21.2} & {2} & {8} & {8} & {8} & {1024} & {Yes} & 1 \\ \hline 
        {20.5} & {2} & {8} & {8} & {8} & {1024} & {Yes} & No \\ \hline 
        {20.4}  & {2} & {4} & {4}  & {16} & {2048}&  {Yes} & No \\ \hline  % Mirco's best model
        20.2 & 2 & 4 & 4  & 8 &2048 &  Yes & No  \\ \hline % dpt 35 (cedar) 
        %20.2 & 2 & 8 & 8 & 8 & 1024 & Yes \\ \hline  % Jian -- this got to 20.5!
        19.9 & 2 & 4 & 4 & 8 & 2048 & Yes & No \\ \hline % dpt 25
        19.8 & 3 & 4 & 4  & 8 & 2048 & Yes & No \\ \hline  %dpt 26
        19.4 & 2 & 4 & 4 & 8 & 2048 & No & No \\ \hline  %dpt 18
        19.2 & 2 & 4 & 1 & 8 & 2048 & Yes & No\\ \hline %Mirko ablation
        19.1  & 2 & 3 & 3  & 8 &2048 &  Yes & No \\ \hline % dpt22
        19.0  & 2 & 3 & 3  & 8  & 2048 & No & No \\ \hline % dpt15
    \end{tabular}
    }
     \label{tab:ablation}
\end{table}

%\vspace{-0.3cm}
\subsection{Ablation Study} 

Hereafter we study the effect of various hyperparameters and data augmentation on the performance of the SepFormer using WSJ0-2mix dataset. The results are summarized in Table \ref{tab:ablation}. The reported performance in this table is calculated on the validation set.

We observe that the number of InterT and IntraT blocks has an important impact on the performance. The best results are achieved with 8 layers for both blocks replicated two times. We also would like to point out that a respectable performance of 19.2 dB is obtained even when we use a single layer transformer for the InterTransformer. This suggests that the IntraTransformer, and thus local processing, has a greater influence on the performance. It also emerges that positional encoding is helpful (e.g. see lines 3 and 5 of Table \ref{tab:ablation}). A similar outcome has been observed in \cite{kim2020t} for speech enhancement. As for the number of attention heads, we observe a slight performance difference between 8 and 16 heads. Finally, it can be observed that dynamic mixing helps the performance significantly. 

\vspace{-0.25cm}
\subsection{Results on WSJ0-3mix}
\vspace{-0.1cm}
Table \ref{table:3mix} showcases the best performing models on the WSJ0-3mix dataset. SepFormer obtains the state-of-the-art performance with an SI-SNRi of 19.5 dB and an SDRi of 19.7 dB.
We used here the best architecture found for the WSJ0-2mix dataset. The only difference is that the decoder has now three outputs. 
It is worth noting that on this corpus the SepFormer outperforms all previously proposed systems. %In particular, the SepFormer outperforms Wavesplit \cite{zeghidour2020wavesplit} even though our model does not use speaker identities. 

%To deal with the additional complexity of this dataset, we used a larger model with 26M parameters which employs $\numinter = 8$ Inter transformer layers, $\numintra=8$ Intra transformer layers, and we use $N=2$. We use 8 parallel attention heads and 1024 dimensions for the feedforward layers in the transformer layers. We again used a stride factor of 8 in the convolutional encoder. %We call this model Sepformer Deep in Figure \ref{fig:forwardpass} (left-middle).

Our results on WSJ0-2mix and WSJ0-3mix show that it is possible to achieve state-of-the-art performance in separation with an RNN-free Transformer-based model. The big advantage of SepFormer over RNN-based systems like \cite{luo2020dualpath, nachmani2020voice, dptn} is the possibility to parallelize the computations over different time steps. This leads to faster training and inference, as described in the following section.

 \begin{table}[t]
 \caption{Best results on the WSJ0-3mix dataset.}
 \vspace{0.1cm}
 \label{table:3mix}
 \centering
 \resizebox{7.6cm}{!}{
\begin{tabular}{l|c|c|c}
\textbf{Model} & \textbf{SI-SNRi} & \textbf{SDRi} & \textbf{\# Param} \\
\hline \hline 
ConvTasnet \cite{luo2018convtasnet} & 12.7 & 13.1 & 5.1M\\ \hline 
DualPathRNN \cite{luo2020dualpath} & 14.7 & n.a & 2.6M \\ \hline
VSUNOS \cite{nachmani2020voice} & 16.9 & n.a & 7.5M  \\ \hline
Wavesplit \cite{zeghidour2020wavesplit} & 17.3 & 17.6 & 29M \\ 
Wavesplit \cite{zeghidour2020wavesplit} + DM & 17.8 & 18.1 & 29M \\ \hline \hline
\textbf{Sepformer} &  17.6 & 17.9 & 26M \\   
\textbf{Sepformer + DM} &  19.5 & 19.7 & 26M \\ \hline  
\end{tabular}
}
\end{table}

\vspace{-0.40cm}
\subsection{Speed and Memory Comparison}
\label{sec:speed}

We now compare the training and inference speed of our model with DPRNN \cite{luo2020dualpath} and DPTNet \cite{dptn}. Figure \ref{fig:forwardpass} (left) shows the training curves of the aforementioned models on the WSJ0-2mix dataset. We plot the performance achieved on the validation set in the first 48 hours of training versus the wall-clock time. For a fair comparison, we used the same machine with the same GPU (a single NVIDIA V100-32GB) for all the models. Moreover, all the systems are trained with a batch size of 1 and employ automatic mixed precision.  We observe that the SepFormer is faster than DPRNN and DPTNeT.  Figure \ref{fig:forwardpass} (left), highlights that SepFormer 
%our model (which achieves a final performance of 20.4 dB on the validation set of WSJ0-2Mix) 
reaches above 17dB levels only after a full day of training, whereas the DPRNN model requires two days of training to achieve the same level of performance.  

Figure \ref{fig:forwardpass} (middle\&right) compares the average computation time (in ms) and the total memory allocation (in GB) during inference when single precision is used. We analyze the speed of our best model for both WSJ0-2Mix and WSJ0-3Mix datasets. We compare our models against DP-RNN, DPTNeT, and Wavesplit. All the models are stored in the same NVIDIA RTX8000-48GB GPU and we performed this analysis using the PyTorch profiler \cite{pytorch-profiler}. For Wavesplit we used the implementation in \cite{pariente:hal-02962964}.

From this analysis, it emerges that the SepFormer is not only faster but also less memory demanding than DPTNet, DPRNN, and Wavesplit. We observed the same behavior using the CPU for inference also. Such a level of computational efficiency is achieved even though the proposed SepFormer employs more parameters than the other RNN-based methods (see Table \ref{table:WSJ2Mix}).
This is not only due to the superior parallelization capabilities of the proposed model, but also because the best performance is achieved with a stride factor of 8 samples, against a stride of 1 for DPRNN and DPTNet. Increasing the stride of the encoder results in downsampling the input sequence, and therefore the model processes less data. In \cite{luo2020dualpath}, the authors showed that the DPRNN performance degrades when increasing the stride factor. %, although this results in a faster model. 
The  SepFormer, instead, reaches competitive results even with a relatively large stride, leading to the aforementioned speed and memory advantages.

\vspace{-0.3cm}
\section{Conclusions}
\vspace{-0.2cm}
In this paper, we proposed a novel neural model for speech separation called SepFormer (Separation Transformer).
The SepFormer is an RNN-free architecture that employs a masking network composed of transformers only. The masking network learns both short and long-term dependencies using a multi-scale approach. Our results, reported on the WSJ0-2mix and  WSJ0-3mix datasets, highlight that we can reach state-of-the-art performances in source separation without using RNNs in the network design. This way, computations over different time-steps can be parallelized. Moreover, our model achieves a competitive performance even when subsampling the encoded representation by a factor of 8. These two properties lead to a significant speed-up at training/inference time and a drastic reduction of memory usage, especially when compared to recent models such as DPRNN, DPTNet, and Wavesplit. %Our model also achieves state of the art performance on WSJ0-3mix dataset. 
As future work, we would like to explore different transformer architectures that could potentially further improve performance, speed, and memory usage. 
\bibliographystyle{IEEEbib}
\bibliography{refs}

\end{document}